\documentclass[journal]{IEEEtran}

\ifCLASSINFOpdf
\else
   \usepackage[dvips]{graphicx}
\fi
\usepackage{url}

\usepackage{graphicx}
\usepackage{amsmath}
\usepackage{amsfonts}
\usepackage{subcaption}
\usepackage{microtype}
\usepackage[hidelinks]{hyperref}

\begin{document}

\title{Interpreting Speaker Characteristics in the Dimensions of Self-Supervised Speech Features}

\author{Kyle Janse van Rensburg, Benjamin van Niekerk, and Herman Kamper
\thanks{K. Janse van Rensburg and H. Kamper are with the Department of Electrical and Electronic Engineering, Stellenbosch University, South Africa (email: kylejvr767@gmail.com, kamperh@sun.ac.za).}

\thanks{B. van Niekerk was with the Department of Electrical and Electronic Engineering, Stellenbosch University, South Africa, when this work was carried out. He is currently with Concordia University, Canada (email: benjamin.l.van.niekerk@gmail.com).}
}

\markboth{Published in IEEE Signal Processing Letters}
{}
\maketitle

\begin{abstract}

How do speech models trained through self-supervised learning structure their representations?
Previous studies have looked at how information is encoded in feature vectors across different layers.
But few studies have considered whether speech characteristics are captured \textit{within} individual dimensions of SSL features.
In this paper, we specifically look at speaker information using PCA on utterance-averaged representations.
For a range of SSL models, we find that the principal dimension that explains most variance encodes pitch and associated characteristics like gender.
Other individual principal dimensions correlate with intensity, noise levels, the second formant, and higher frequency characteristics.
We then use synthesis analyses to show that the dimensions for most characteristics are isolated from each other's influence.
We further show that characteristics can be changed by manipulating the corresponding dimensions.
\end{abstract}


\IEEEpeerreviewmaketitle

\section{Introduction}
\label{sec:intro}

\IEEEPARstart{S}{elf}-supervised learned (SSL) speech models have become integral in speech processing systems, producing rich speech features that encode both linguistic and speaker information~\cite{SUPERB}.
But how do these models structure the information in their representations?
And can we take advantage of this structure, for instance, for performing simple manipulations to change a voice in an SSL-based synthesis system?

Multiple studies have analysed the information encoded in SSL features. 
Large-scale layer-wise analyses have been performed by training classifiers such as linear models~\cite{Chung_etal} or feedforward networks~\cite{Chiu_etal} on top of the SSL representations to see how accessible content and speaker information are.
Pasad et al.~\cite{pasad2021,pasad2023} performed probing experiments directly on the representations of different SSL models' layers using canonical correlation analysis to measure how and in which layers different properties are captured.
From these studies, there seems to be a hierarchy of information encoded in the layers, with speaking attributes and finer-grained phonetic information encoded in the initial layers, while higher-level information like lexical properties are correlated with later layers.
Furthermore, the specific identity markers captured by these types of representations reveal a strong bias toward static features over dynamic, behavioural speech patterns~\cite{carbonneau2025}.

While previous work provide a perspective on how information is encoded across layers and across models, few studies have considered whether speech characteristics are captured within individual dimensions of SSL features.
Liu et al.~\cite{Liu2023} ventured into this space by applying principal component analysis (PCA) to self-supervised predictive coding models and showed that speaker and phonetic content are captured in orthogonal subspaces.
Gubian et al.~\cite{Gubian2025} showed that this is true even if the language used during pretraining is different from the test language.
Both of these studies, however, focused on phone-level representations, relying on time-aligned boundaries to analyse the SSL features.
We are instead interested in how speaker information is encoded within the SSL space.

In this work, we directly examine the individual dimensions of SSL representations to determine the structure of how speaker characteristics are encoded.
We specifically look at speaker information using PCA on utterance-level averaged SSL features obtained from Wav2Vec2~\cite{wav2vec2}, HuBERT~\cite{HuBERT}, and WavLM~\cite{WavLM}.
Correlation analysis is applied between individual principal dimensions and specific speaker characteristics, such as pitch, intensity, and timbre.
We find that, overall, WavLM produces the best correlation results, and therefore perform further analysis on this representative model.

We find that a single dimension jointly encodes several speaker characteristics such as pitch, gender, and local jitter, a measurement of the stability of the pitch in an utterance.
This dimension is also the one with the highest variance in the PCA projection.
Other individual dimensions correlate with intensity, noise levels, higher-order formants, and the presence of higher frequency content and noise.
Do the dimensions capture the characteristics in a decoupled manner?
To answer this question, we use synthesis experiments: we alter specific speaker characteristics by changing the corresponding principal dimension and then synthesise the result with a vocoder~\cite{HiFi-GAN}.
For many characteristics, we find that alterations can be made by simply changing the corresponding dimensions.
Importantly, we find that control is largely isolated: changing one dimension does not affect non-correlated speaker characteristics.

This study complements existing work on understanding the geometry of SSL speech features, showing specifically how speaker characteristics are captured in specific directions.

\section{Methodology}
\label{sec:methodology}

Our goal is to analyse the relationship between speaker characteristics and SSL features to determine how the SSL space is structured.
We are specifically interested in whether individual speaker characteristics are encoded in particular directions.
We therefore measure the correlation between specific characteristics and the features' projections onto individual principal directions.

\subsection{Speaker characteristics}
\label{subsec:speak_charac}

We consider the following speaker-specific characteristics: pitch averaged over an utterance (F0, Hz); F1, F2, and F3 formant values averaged over an utterance (Hz); intensity (dB); local jitter (\%); local shimmer (\%); speaking rate (phones/sec); harmonic-to-noise ratio (HNR, dB); the spectral rolloff point~(Hz); zero-crossing rate (ZCR); and the speaker's gender. 
Jitter gives an indication of the stability of pitch over an utterance, while shimmer indicates variation in intensity~\cite{jitter_shimmer}.
HNR can be seen as a measure of noise levels~\cite{hnr}, while ZCR gives an indication of both voicing and noisiness; both these characteristics are also affected by the channel's properties~\cite{zcr_spec_roll}.
The spectral rolloff point gives a measure of the relative energy found in high vs low frequencies~\cite{scheirer_etal,zcr_spec_roll}. It is, in our case, the frequency below which 50\% of the signal's energy resides, i.e., higher values indicate more content in higher frequencies.

Pitch, formants, intensity, jitter, shimmer, and HNR are calculated using Parselmouth's Praat functions~\cite{praat,parselmouth}.
Speaking rate is calculated using the forced alignments of the data.
ZCR and spectral rolloff are calculated using Librosa~\cite{librosa}.

\subsection{Principal component analysis on SSL features}
\label{subsec:pca_on_SSL}

PCA allows us to find the orthogonal principal directions within the SSL feature space that capture the most variance.
We can then project the data onto these principal directions, producing the principal dimensions.
In our case, each data point consists of SSL features averaged over an utterance from a single speaker. 
We want to know whether and how the principal dimensions of these features capture speaker information.
PCA will not capture non-linear structure in the space; it therefore provides a way to determine which characteristics are most salient.

For training the PCA models, we use the full LibriSpeech train-clean-100 dataset~\cite{LibriSpeech} to ensure sufficient data for stable component estimation and to reduce sensitivity to recording noise. 
We use scikit-learn's PCA implementation, with 50 principal components~\cite{scikit-learn}.

To subsequently assess the correlation between speaker characteristics and individual principal dimensions, we need speech audio that is of sufficient quality that speech characteristics can be accurately extracted.
However, during data preprocessing of the LibriSpeech datasets, we noticed qualitatively that the noise in some audio affected the accuracy of the extraction of the speaker characteristics.
We therefore develop smaller, curated datasets of utterances to ensure accurate feature extraction to be used as ground-truth measurements for the correlation analysis.
We manually selected 10 utterances from 10 speakers (5 male and 5 female) from the dev-clean and test-clean sets, respectively, by viewing some of the speaker characteristics (like pitch) on the audio spectrograms.
The result is curated validation and test sets of 100 utterances each.

\begin{figure}[!t]
    \centerline{\includegraphics[width=\columnwidth]{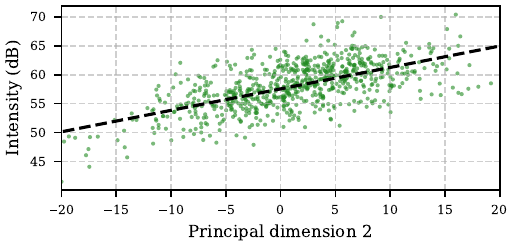}}
    \caption{The relationship between principal dimension 2 and intensity on a subset of train-clean-100 data, with $R^2 = 0.40$.}
\label{fig:intensity_linear}
\end{figure}

\begin{figure*}[t!]
    \centerline{\includegraphics[width=\textwidth]{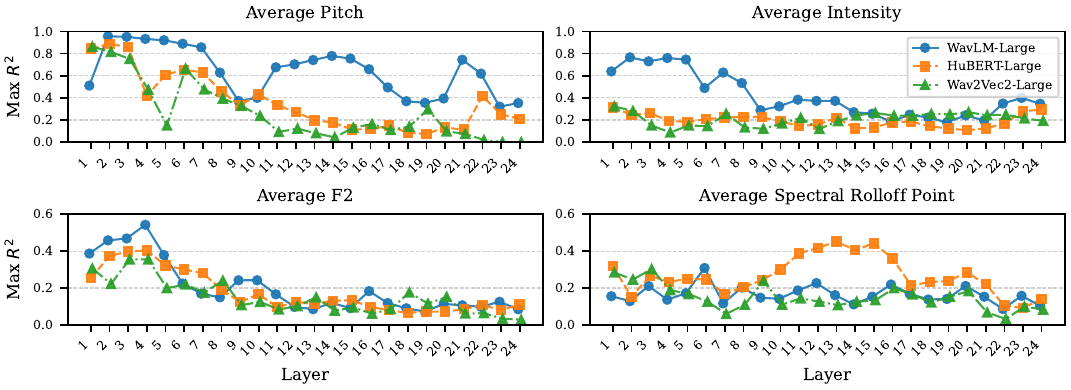}}
    \caption{The top correlation scores for each layer of each SSL model for average pitch, intensity, F2, and the spectral rolloff point.}
    \label{fig:layer_analysis}
\end{figure*}

\subsection{Correlation analysis}
\label{subsec:corr_analysis}

To determine whether a particular dimension tells us something about a particular speaker characteristic, we use correlation analysis.
All our speaker characteristics are continuous except for gender, which is treated as a categorical variable.

To measure correlation for continuous characteristics, we use the coefficient of determination, $R^2$, a score in the range of $(-\infty, 1]$~\cite{ISL_python}.
A score of 1 indicates a perfect linear relationship between the speaker characteristic and the dimension under investigation.
A score of 0 indicates that the prediction from the dimension is as bad as consistently predicting the data mean.
A negative score indicates the correlation is worse than predicting the mean.
To give an intuitive understanding, Fig.~\ref{fig:intensity_linear} shows the relationship between the principal dimension 2 compared to utterance intensity.

For the categorical gender label, we use Cohen’s kappa ($\kappa$), a score in the range of $[-1, 1]$~\cite{cohen_kappa}.
A 1 indicates perfect agreement between the gender label and a binary classification using the dimension value.
A 0 is equivalent to the model predicting a random gender label, while a negative score indicates the model's performance is worse than random.

\section{Layer-Wise Correlation Analysis}
\label{sec:layer_analysis}

We start with a layer-wise analysis of a range of SSL models.
Our goal is to identify a representative model to use in our in-depth dimension-wise analysis (Section \ref{sec:analysis_princ_dim}).

We consider the Wav2Vec2~\cite{wav2vec2}, HuBERT~\cite{HuBERT}, and WavLM~\cite{WavLM} SSL models.
For each layer from each model, correlation is calculated between each principal dimension and every speaker characteristic. We then note the top correlation score for that particular layer.
Fig. \ref{fig:layer_analysis} shows these correlation results for a subset of the characteristics.

For a majority of the characteristics, the models achieve the best correlations in the initial layers (1 to 10), with scores then dropping off in later layers.
There are characteristics such as pitch where the correlations rise and fall for WavLM, and the spectral rolloff point where HuBERT achieves the highest correlations in the middle layers, but these are outliers.
Our analysis shows that finer phonetic and speaker-specific information is captured in the initial layers.
Of all the models, WavLM achieves the highest correlation scores for the majority of characteristics. 
We therefore use this model for our per-dimension analysis below.

\section{Analysis of Principal Dimensions}
\label{sec:analysis_princ_dim}

\begin{figure}
    \centerline{\includegraphics[width=\columnwidth]{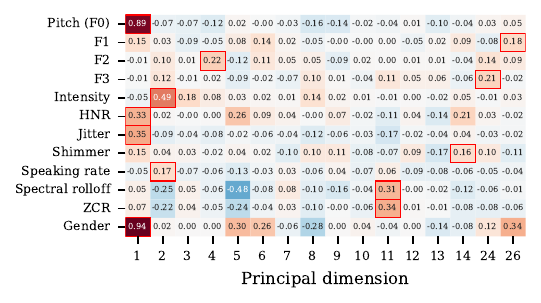}}
    \caption{Heat map showing correlation scores between speaker-specific characteristics and principal dimensions for WavLM-Large layer 6. Development data is used.}
\label{fig:wavlm_6_correlations}
\end{figure}

In this section, we analyse the correlations between principal dimensions and different speaker characteristics of WavLM's initial layers.
The goal is to determine if a particular dimension captures one (or more) of the speaker characteristics.

Fig.~\ref{fig:wavlm_6_correlations} shows the $R^2$ and $\kappa$ scores for the correlation analysis for each speaker-specific characteristic of WavLM's layer 6 on the development dataset.
For conciseness, only a subset of the principal dimensions with the highest correlations are shown.
The principal dimension that scores the highest for each characteristic is indicated with a red outline.

Principal dimension 1 is the top correlated dimension for average pitch and gender.
Since females generally have a higher pitch than males, it makes sense that these characteristics are associated with a single dimension.
Furthermore, dimension~1 shows the strongest correlation with HNR and jitter, as well as the second-strongest correlation with the average F1 and shimmer.
These characteristics are associated with timbre, which indicates that dimension 1 also captures coarser properties of a speaker's speaking style.
HNR, however, is also associated at a similar strength across dimensions 1, 5, and 14.

Principal dimension 2 is correlated with the average intensity and speaking rate of an utterance.
Dimension 4 is correlated with the average F2 while dimension 11 captures the voicing of an utterance, specifically its spectral energies in the spectral rolloff and the amount of voiced content in the ZCR.
Principal dimension 14 captures shimmer, dimension 24 captures  F3, and dimension 26 is associated with F1.

To see whether these results generalise, we also performed the same analysis on WavLM's layers 5 and 7. These aren't shown, but are very similar to Fig.~\ref{fig:wavlm_6_correlations}.
Layer 6, however, does generally achieve the highest correlation scores, and has more individual dimensions that are only associated with one or two speaker characteristics.

Taking our analyses together, it seems that the first principal direction with the most variance captures coarse speaker-specific characteristics, the second direction captures intensity and speaking rate, and subsequent directions are often associated with single characteristics.
A number of studies have employed PCA as a dimensionality reduction approach for SSL features~\cite{speak_verif_pca,simon_2025,danel_2026}, and the analysis here reveals the main speaker characteristics retained when doing dimensionality reduction on averaged SSL features.

We have shown that correlation exists; next, we look at whether any of the dimensions influence each other in synthesis experiments.

\section{Analysis via Synthesis}
\label{sec:synthesis}

We want to know whether the principal dimensions associated with specific speaker characteristics are uncoupled, i.e., when a dimension is changed, does it only influence its correlated characteristic(s), and no other?
We do this via synthesis experiments, where we modify the principal dimension and then track how speaker characteristics are altered after vocoding the modified SSL feature sequence.

\subsection{Experimental setup}
\label{subsec:synthesis_exp_setup}

Concretely, we take an utterance and modify one of its principal dimensions.
We then resynthesise the audio and measure the speaker characteristics.
We analyse how the characteristics change as we modify the dimensions, determining their degrees of influence.

Formally, given an SSL feature sequence $\mathbf{x}_1, \mathbf{x}_2, \ldots \mathbf{x}_T$ from an utterance, we modify a particular dimension $i$ by adding a scalar multiple of the unit length principal direction $\mathbf{v}_i$ to each frame, such that $\mathbf{x}_{n}^{\text{mod}} = \mathbf{x}_n + \alpha \sigma_{i} \cdot \mathbf{v}_i$.
Here, $\sigma_{i}$ is the standard deviation for principal dimension $i$.\footnote{This equation holds for zero-mean data; in practice the data bias is also accounted for.}
We then synthesise the speech from the modified frames using a pretrained HiFi-GAN model~\cite{HiFi-GAN,kNN-VC}.
This is the second reason we chose WavLM layer 6 in Section~\ref{sec:analysis_princ_dim}, as the available vocoder is specifically trained on layer 6 representations.

\subsection{Results}
\label{subsec:synthesis_results}

\begin{figure}
    \centerline{\includegraphics[width=\columnwidth]{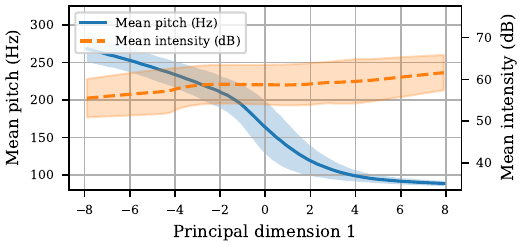}}
    \caption{An illustration of how varying a principal dimension affects other
    characteristics than the one that it is correlated with. Here, principal dimension 1 (associated with pitch) is varied, and intensity is also measured. The blue solid line shows the average pitch and the dashed orange line shows the average intensity over all utterances in the test set as the dimension changes. The shaded areas indicate one standard deviation from the mean.}
\label{fig:control_pitch_intensity}
\end{figure}

We start our analysis by attempting to quantify the degree of control which correlated dimensions have over their characteristics.
Fig.~\ref{fig:control_pitch_intensity} shows the synthesis analysis results when varying dimension 1 (associated with pitch and gender) to change average pitch, with the blue line indicating the average pitch from all test utterances after performing the modification.
The shaded blue area shows one standard deviation of the measured pitch.
When qualitatively listening to the manipulated audio, we hear that it also maintains good audio quality (we provide a demo webpage with audio samples).\footnote{\url{https://kylejvr767.github.io/SPL_demo_page/}}
We see that we can change the average utterance pitch to be between 100 and 300~Hz.
To test the effect on individual utterances, we also selected naturally high- and low-pitched test utterances and varied them (not pictured).
We find that they follow the same trend as the average, with their pitch values within about two standard deviations.
But there are limits: pitch values start to plateau as principal dimension 1 is moved excessively far from the data distribution.
In Fig.~\ref{fig:control_pitch_intensity} we see this occur
when we move beyond 4 and $-$4 standard deviations from the mean.
Low-pitched utterances plateau very quickly when the utterance is being pitched down, and vice versa for high-pitched utterances.

Expanding these analyses to every speaker characteristic, we find that pitch is the only characteristic which exhibits a plateau; intensity, F2, F3, the spectral rolloff point, and ZCR can be manipulated in a linear fashion.
The remaining characteristics (F1, jitter, shimmer, and HNR) did not change systematically when their corresponding dimension was altered.

Next, we test the principal dimensions' influence leakage.
We repeat the same analyses for the correlated dimensions, but now we measure the effect the manipulation has on all of the uncorrelated characteristics.
Starting again with dimension 1, Fig.~\ref{fig:control_pitch_intensity} shows that as pitch is changed when varying dimension 1, the average intensity (orange line) remains relatively stable throughout.
Changing the pitch to be between 100 and 300~Hz results only in a 5 dB change in intensity.
As reference, changing dimension 2 (associated with intensity) can change the mean intensity to be anywhere between 40 and 70~dB.
Similar effects are seen when measuring the other characteristics that are not correlated with dimension 1 (not shown here).
Moreover, the same relationship is observed when varying the other correlated principal directions and measuring their correlated vs. uncorrelated characteristics.
This shows that when changing a principal dimension, the change is isolated.

This analysis reveals the potential of using our approach as an immediate, simple, training-free method to manipulate the output of a speaker's voice via the SSL representations in a voice conversion system.
This adds to a growing body of work~\cite{kNN-VC,LinearVC,ZeroSyl} that shows that direct manipulation of SSL representations can be used to solve some speech processing tasks without resorting to large, intricate model training.

\section{Conclusion}
\label{sec:conclusion}

This paper considered how the Wav2Vec2, HuBERT, and WavLM SSL models encode speaker characteristics in particular directions of their resulting features.
To understand how speaker characteristics are encoded, we used PCA to find the principal dimensions that capture most of the variance for features averaged over utterances.
We found that the first dimension captured pitch and gender as well as coarser speaker characteristics like jitter; subsequent dimensions were mostly correlated with only one or two speaker characteristics.
In synthesis experiments, we showed that the various correlated principal dimensions are uncoupled from each other. By varying the dimensions individually, we could change the corresponding speaker characteristic, at least to some degree: features like pitch, intensity, and some higher-order formants could be changed, but features like jitter and shimmer could not.

Future work may expand on our synthesis experiments to include other SSL models and hidden layers to determine which allows for the greatest degree of precision for changing speaker characteristics.
Further analysis should also be conducted to determine why some characteristics can be changed while others cannot.
Our work has many potential applications in areas such as voice conversion, fictional character generation, and voice anonymisation~\cite{knn_vc_anonymisation,child_anonymisation_knn_vc}.
Finally, applications in voice anonymisation should be studied with these techniques to determine if changing the principal dimensions are enough to obfuscate speaker identity.


\begin{thebibliography}{34}

\bibitem{SUPERB}
    S. Yang,
    P.-H. Chi,
    Y.-S. Chuang,
    C.-I. J. Lai,
    K. Lakhotia,
    Y. Y. Lin,
    A. T. Liu,
    J. Shi,
    X. Chang,
    G.-T. Lin,
    T.-H. Huang,
    W.-C. Tseng,
    K.-T. Lee,
    D.-R. Liu,
    Z. Huang,
    S. Dong,
    S.-W. Li,
    S. Watanabe,
    A. Mohamed,
    H.-Y. Lee,
    ``SUPERB: Speech processing universal performance benchmark,''
    in \textit{Proc. Interspeech},
    2021.

\bibitem{Chung_etal}
    Y. -A. Chung,
    Y. Belinkov,
    J. Glass,
    ``Similarity analysis of self-supervised speech representations,''
    in \textit{Proc. ICASSP},
    2021.

\bibitem{Chiu_etal}
    A. Y. F. Chiu,
    K. C. Fung,
    R. T. Y. Li,
    J. Li,
    T. Lee,
    ``A large-scale probing analysis of speaker-specific attributes in self-supervised speech representations,''
    \textit{arXiv preprint arXiv:2501.05310},
    2025.

\bibitem{pasad2021}
    A. Pasad,
    J.-C. Chou,
    K. Livescu,
    ``Layer-wise analysis of a self-supervised speech representation model,''
    in \textit{Proc. ASRU},
    2021.

\bibitem{pasad2023}
    A. Pasad,
    B. Shi,
    K. Livescu,
    ``Comparative layer-wise analysis of self-supervised speech models,''
    in \textit{Proc. ICASSP},
    2023.

\bibitem{carbonneau2025}
    M.-A. Carbonneau, 
    B. van Niekerk,
    H. Seuté,
    J.-P. Letendre,
    H. Kamper,
    J. Zaïdi,
    ``Analyzing and improving speaker similarity assessment for speech synthesis,''
    in \textit{Proc. SSW},
    2025.

\bibitem{Liu2023}
    O. D. Liu,
    H. Tang,
    S. Goldwater,
    ``Self-supervised predictive coding models encode speaker and phonetic information in orthogonal subspaces,''
    in \textit{Proc. Interspeech},
    2023.

\bibitem{Gubian2025}
    M. Gubian,
    I. Krehan,
    O. Liu,
    J. Kirby,
    S. Goldwater,
    ``Analyzing the relationships between pretraining language, phonetic, tonal, and speaker information in self-supervised speech models,''
    \textit{arXiv preprint arXiv:2506.10855},
    2025.

\bibitem{wav2vec2}
    A. Baevski,
    H. Zhou,
    A. Mohamed,
    M. Auli,
    "Wav2Vec 2.0: A framework for self-supervised learning of speech representations", 
    in \textit{Proc. NeurIPS},
    2020.

\bibitem{HuBERT}
    W. -N. Hsu,
    B. Bolte,
    Y. H. H. Tsai,
    K. Lakhotia,
    R. Salakhutdinov,
    A. Mohamed,
    "HuBERT: Self-supervised speech representation learning by masked prediction of hidden units", 
    in \textit{IEEE/ACM Transactions on Audio, Speech, and Language Processing},
    2021.

\bibitem{WavLM}
    S. Chen,
    C. Wang,
    Z. Chen,
    Y. Wu,
    S. Liu,
    Z. Chen,
    J. Li,
    N. Kanda,
    T. Yoshioka,
    X. Xiao,
    J. Wu,
    L. Zhou,
    S. Ren,
    Y. Qian,
    Y. Qian,
    M. Zeng,
    X. Yu,
    F. Wei,
    ``WavLM: large-scale self-supervised pre-training for full stack speech processing,''
    \textit{IEEE Journal of Selected Topics in Signal Processing},
    2022.

\bibitem{HiFi-GAN}
    J. Kong,
    J. Kim,
    J. Bae,
    ``HiFi-GAN: generative adversarial networks for efficient and high fidelity speech synthesis,''
    in \textit{Proc. NeurIPS},
    2020.

\bibitem{kNN-VC}
    M. Baas,
    B. van Niekerk,
    H. Kamper,
    ``Voice conversion with just nearest neighbors,''
    in \textit{Proc. Interspeech},
    2023.

\bibitem{LinearVC}
    H. Kamper, 
    B. van Niekerk,
    J. Zaïdi,
    M.-A. Carbonneau,
    ``LinearVC: Linear transformations of self-supervised features through the lens of voice conversion,''
    in \textit{Proc. Interspeech},
    2025.

\bibitem{ZeroSyl}
    N. Visser,
    S. Malan,
    D. Slabbert,
    H. Kamper,
    ``ZeroSyl: Simple zero-resource syllable tokenization for spoken language modeling,''
    in \textit{Proc. Interspeech},
    2026.

\bibitem{jitter_shimmer}
    M. Farrús,
    J. Hernando,
    P. Ejarque,
    ``Jitter and shimmer measurements for speaker recognition,''
    in \textit{Proc. Interspeech},
    2007.

\bibitem{hnr}
    F. Jalali-najafabadi,
    C. Gadepalli,
    D. Jarchi,
    B. M. G. Cheetham,
    ``Acoustic analysis and digital signal processing for the assessment of voice quality,''
    \textit{Biomedical Signal Processing and Control},
    2021.

\bibitem{zcr_spec_roll}
    F. Alías,
    J. C. Socoró,
    X. Sevillano,
    ``A review of physical and perceptual feature extraction techniques for speech, music and environmental sounds,''
    \textit{Applied Sciences},
    2016.
    
\bibitem{scheirer_etal}
    E. Schreirer,
    M. Slaney,
    ``Construction and evaluation of a robust multifeature speech/music discriminator,''
    in \textit{Proc. ICASSP},
    1997.

    \bibitem{praat}
    P. Boersma,
    ``Praat, a system for doing phonetics by computer,''
    \textit{Glot International},
    2001.

\bibitem{parselmouth}
    Y. Jadoul,
    B. Thompson,
    B. de Boer,
    ``Introducing Parselmouth: A Python interface to Praat,''
    \textit{Journal of Phonetics},
    2018.

\bibitem{librosa}
    B. McFee,
    C. Raffel,
    D. Liang,
    \textit{et al.},
    ``Librosa: Audio and music signal analysis in Python,''
    in \textit{Proc. SciPy},
    2015.

\bibitem{LibriSpeech}
    V. Panayotov, 
    G. Chen, 
    D. Povey, 
    S. Khudanpur, 
    ``Librispeech: An ASR corpus based on public domain audio books,'' 
    in \textit{Proc. ICASSP},
    2015.

\bibitem{scikit-learn}
    F. Pedregosa,
    G. Varoquaux,
    A. Gramfort,
    \textit{et al.},
    ``Scikit-learn: Machine learning in Python,''
    \textit{Journal of Machine Learning Research},
    2011.

\bibitem{ISL_python}
    G. James,
    D. Witten,
    T. Hastie,
    R. Tibshirani,
    J. Taylor,
    \textit{An Introduction to Statistical Learning with Applications in Python},   
    Springer,
    1st ed,
    2023.

\bibitem{cohen_kappa}
    J. Cohen,
    ``A coefficient of agreement for nominal scales,''
    \textit{Educational and Psychological Measurement},
    1960.

\bibitem{speak_verif_pca}
    W.-H. Liao,
    P.-H. Chen,
    Y.-C. Wu,
    ``Unveiling the potential of SSL-generated audio embeddings for cross-lingual speaker recognition,''
    in \textit{Proc. ISM},
    2024.

\bibitem{simon_2025}
    S. Malan,
    B. van Niekerk,
    H. Kamper,
    ``Should top-down clustering affect boundaries in unsupervised word discovery,''
    \textit{IEEE/ACM Transactions on Audio, Speech and Language Processing},
    2026.

\bibitem{danel_2026}
    D. Slabbert,
    S. Malan,
    H. Kamper,
    ``Unsupervised lexicon learning from speech is limited by representations rather than clustering,''
    in \textit{Proc. ICASSP},
    2026.

\bibitem{knn_vc_anonymisation}
    A. Das,
    C. Franzreb,
    T. Herzig,
    P. Pirlet,
    T. Polzehl,
    ``Comparing speech anonymization efficacy by voice conversion using KNN and disentangled speaker feature representations,''
    in \textit{Proc. SPSC},
    2024.

\bibitem{child_anonymisation_knn_vc}
    A. Kulkarni,
    F. Teixeira,
    E. Hermann,
    T. Rolland,
    I. Trancoso,
    M. Magimai-Doss,
    ``Children’s voice privacy: first steps and emerging challenges,''
    in \textit{Proc. Interspeech},
    2025.

\end{thebibliography}
\end{document}